\documentclass[aip,graphicx]{revtex4-1}
\usepackage{graphicx}% Include figure files
\usepackage{dcolumn}% Align table columns on decimal point
\usepackage{bm}% bold math
\usepackage[usenames,dvipsnames]{xcolor}
\usepackage[utf8]{inputenc}
\usepackage[T1]{fontenc}
\usepackage{mathptmx}
\usepackage{hyperref}

\newcommand{\ld}{\lambda_\mathrm{D}}

\newcommand{\kbt}{k_\mathrm{B}T}

\newcommand{\tc}{\tau_c}
\newcommand{\td}{\tau_D}

\draft % marks overfull lines with a black rule on the right
\makeatletter
\let\@fnsymbol\@fnsymbol@latex
\@booleanfalse\altaffilletter@sw
\makeatother

\begin{document}

%\preprint{}

\title{Computing the Local Ion Concentration Variations for Electric-Double-Layer-Modulation Microscopy} %Title of paper



\author{Zhu Zhang}
\affiliation{Nanophotonics, Debye Institute for Nanomaterials Science, Utrecht University, 3584CC Utrecht, The Netherlands}
\author{Jie Yang}
\affiliation{School of Chemistry and Molecular Engineering, East China University of Science and Technology, Shanghai 200237, China}
\author{Cheng Lian}
\email[]{liancheng@ecust.edu.cn}
\affiliation{School of Chemistry and Molecular Engineering, East China University of Science and Technology, Shanghai 200237, China}
\author{Sanli Faez}
\email[]{s.faez@uu.nl}
\affiliation{Nanophotonics, Debye Institute for Nanomaterials Science, Utrecht University, 3584CC Utrecht, The Netherlands}

\date{\today}

\begin{abstract}
Modulating the electric potential on a conducting electrode is presented to generate an optical contrast for scattering microscopy that is sensitive to both surface charge and local topography. 
We dub this method Electric-Double-Layer-Modulation microscopy. 
We numerically compute the change in the local ion concentration that is the origin of this optical contrast for three experimentally relevant geometries: nanosphere, nanowire, and nanohole. 
In absence of plasmonic effects and physical absorption, the observable optical contrast is proportional to the derivative of the ion concentration with respect to the modulated potential. 
We demonstrate that this derivative depends on the size of the object and, less intuitively, also on its surface charge. 
This dependence is key to measuring the surface charge, in an absolute way, using this method. 
Our results help to identify the experimental conditions such as dynamic range and sensitivity that will be necessary for detecting the elementary charge jumps. 
We conclude that the nanohole is the most suitable geometry among these three for achieving elementary charge sensitivity.
\end{abstract}

\pacs{}% insert suggested PACS : Electrochemistry, electrostatic double layer forces, metrology, liquid-solid interfaces, nanostructures, dark-field microscopy, optical nanoscopy

\maketitle %\maketitle must follow title, authors, abstract and \pacs
\section{Introduction}
%\label{}
%% Importance of EDL
The electric double layer (EDL) is an omnipresent structure at the interface between charged objects and electrolyte solutions:
the solvated ions rearrange to screen the electric field of the charged surface by forming an ionic cloud of opposite charge, typically of a few nanometers thickness. 
Water has an enormous capacity to dissolve ions, because of its large dielectric constant $\epsilon_w \approx 78$ and capacity for ion hydration.
Therefore, from proteins to minerals, many substances obtain a charged surface in contact with water, and hence are surrounded by an EDL. 
The storage and recovery of energy in batteries, the solvation of molecules, the stability of colloids, the filtration process in membranes or in the kidney, and most of mass transport processes in aqueous environments cannot be understood without modelling the EDL~\cite{lyklemaElectricDoubleLayers1995}.
Because of the ubiquitous role of EDL in biological, soft matter, and colloidal systems, experimental research on EDL-related phenomena is still expanding, even a century after its conception~\cite{smithForcesSolidSurfaces2020,haverkortModelingExperimentsBinary2020,faucherCriticalKnowledgeGaps2019, lutzenkirchenSetupSimultaneousMeasurement2018, hashemiamreiOscillatingElectricFields2018,guptaChargingDynamicsOverlapping2020}. 
Recent advances in interferometric scattering microscopy have pushed the sensitivity limit of this method close to the atomic masses, for example a few nucleo-acid blocks~\cite{liSingleMoleculeMass2020a}. 
This sensitivity seems to be limited mainly by two factors: 1- the dynamic range of the detector and 2- the stability of the background scattering ~\cite{taylorInterferometricScatteringMicroscopy2019, youngInterferometricScatteringMicroscopy2019a}.  
As these technical limitations are overcome, illustrated by the other articles in this special issue, 
it seems now feasible to explore detecting small molecules or even single ionisation states. 
At the current pace of increasing sensitivity, even the changes in the refractive index of the investigated particles or their surrounding could soon be detectable on the fly, i.e. without post-detection signal averaging. 
In parallel, other methods such as photothermal microscopy have used the periodicity in modulation of the refractive index as an alternative route to separating signal from a target object from background scattering via lock-in   detection~\cite{adhikariPhotothermalMicroscopyImaging2020, spaethCircularDichroismMeasurement2019}.

Recently, our group demonstrated a new contrast mechanism based on periodic modulation of the surface electrochemical potential~\cite{naminkElectricDoubleLayerModulationMicroscopy2020}.
In this method, modulating the EDL close to the surface results in an scattering signal that is sensitive to both the local topography and the electrochemical properties of the investigated region.
Adding the electrochemical sensitivity to the range of observable quantities for optical phase-sensitive microscopy can expand the already vast range of applications of this technique on several fronts.

In our previous proof of concept report~\cite{naminkElectricDoubleLayerModulationMicroscopy2020}, we have used a simple model to estimate the expected signal from the modulation of EDL.
Our estimated value roughly matches the experimentally observed change in the scattering signal as the surface potential is altered.
This scattering signal is proportional to the concentration of screening ions in the EDL, which is generally a nonlinear function of the surface potential.  
Widely used analytical solutions of the Poisson-Nernst-Plank (PNP) equations that describe the EDL are only valid for small surface potentials of less than $\kbt/e = 25$~mV, with $k_\mathrm{B}$ the Boltzmann constant, $T$ the temperature, and $e$ the unit of charge.
A sufficiently accurate estimation of the potentiodynamic optical contrast thus needs numerical integration, even for the simplest geometries.
In this article, we estimate the range of expected signals for EDL-modulation microscopy of three experimentally relevant geometries using a modified Poisson-Nernst-Plank model that also accounts for the volume exclusion effects of the ions, and hence can be also useful for potentials $\kbt/e > 25$~mV. 
Based on these calculations, we identify the sensitivity of this signal to the size and static charge on the particle in order to answer the central question of this article:
Is it feasible to detect single charging events of small objects with interferometric EDL-modulation microscopy?

\subsection{A brief history of electroreflectance}
 
Investigating the optical signatures of the EDL at the metal-electrolyte interface dates back to Feinlieb's original observation in 1966~\cite{feinleibElectroreflectanceMetals1966}. 
Electric field modulation of optical reflectivity was used at the time to investigate band structures of semiconductors, as the electric field alters charged career dynamics closed to the interface.
Putting an electrolyte in contact with the sample surface allows for obtaining large electric fields with relatively small applied voltages, because of the formation of a thin EDL.
On the metallic side of the interface, the Thomas-Fermi screening length for penetration of this field into the solid scales inversely with the square of the carrier density.  
For semiconductors, this length can be comparable with the visible light wavelength, and hence a large spectral dependence can be expected.
For good conductors, however, this screening occurs within atomic distances from the surface.
The electroreflectance modulation observed at the metal-electrolyte interfaces were, however, comparable with that of semiconductors; about 0.5\% for the spectral peak.
Feinleib's hypothesized that the restructuring of the EDL on the electrolyte side of the interface could partly explain this unexpectedly high modulation, but he also emphasized that the dependence of the measured modulation on the optical frequency could only be explained by considering the electronic band-structure of the metal.
Going through the literature in a decade after this observation, there seem to be an initial disagreement on the explanation~\cite{prostakElectroreflectanceMetals1967, parsonsBandStructureAssociatedAspectsElectroreflectance1969, buckmanElectroreflectanceChangesDielectric1968, bewickStudiesCathodicAdsorption1970}. 
Prostak and Hansen emphasized that the change of the carrier density on the metallic side of the interface could adequately explain the observed electroreflectance and the contribution of EDL restructuring to the reflectivity must be negligible~\cite{prostakElectroreflectanceMetals1967, hansenElectromodulationOpticalProperties1968}.
Stedman (1968) estimated the contribution of the electrolyte ions on changing the reflectivity using an optical model of the EDL based on the known values for polarizability of the anion and cations and found a reasonable match with the previously reported values~\cite{EffectElectricalDouble1968}.
She also refers to preliminary results from an ellipsometry study of mercury-electrolyte interface that could match the signal size expected from double layer reconfiguration, but we could not trace those results in other publications.
Parsons highlighted the discrepancy between the reported measurements on silver and copper~\cite{parsonsBandStructureAssociatedAspectsElectroreflectance1969}.
Bewick and Tuxford measured a similar effect at the platinum electrode in perchloric acid and attributed their observation to the adsorption of hydrogen on platinum~\cite{bewickStudiesCathodicAdsorption1970}.
They also coined the term "modulated specular reflectance spectroscopy" for this technique.
In the electrochemical window, where the EDL effects are most prevalent and Faradaic currents are negligible, results obtained by Bewick and Tuxford were in agreement with Stedman's model. 
They, however, observed a much large modulations in the potential window that Pt is electrochemically active. 
Based on their observations and also Parsons' argument, they argued that the model of Prostak and Hansen fell short of describing electroreflectance in general.
These early observation were summarized in a first review by McIntyre~\cite{McIntyreElectrochemicalModulationSpectroscopy1973}, who emphasized the complexity of experimentally separating various contributions to electroreflectance of metals. 
The change of the permittivity on the electrolyte side of the interface, due to restructuring of the EDL, chemical adsorption of compounds at the interface, and varying optical constants of the metallic side due to a shift in plasmon resonance frequency and dissipative damping were all are considered to play some role in electroreflectance measurements. 
The dependence of the electroreflectance on the optical frequency is the strongest indicator of the contribution of careers in the metallic side. 
However, that alone cannot explain the observed dependence on reversing the direction of the electric field at the interface or the magnitudes of the effect for Pt or Cu interfaces.

Despite the initial attention and widespread adoption of this method in practice, Feinleib's article is scarcely cited in original research reports passed 1984. 
Meanwhile, a variety of other experimental methods have been developed to study the EDL properties at the nanoscale, including electrochemical scanning probe methods~\cite{bardChemicalImagingSurfaces1991, collinsProbingChargeScreening2014, bentleyNanoscaleElectrochemicalMapping2019}, nonlinear spectroscopy~\cite{debeerSeparatingSurfaceStructure2010,lisLiquidFlowSolid2014}, force sensing~\cite{smithElectrostaticScreeningLength2016, leeScalingAnalysisScreening2017a} and nanoparticle tracking~\cite{volpeInfluenceNoiseForce2010}. 
None of these methods, however, has achieved the ultimate single-molecule and elementary charge sensitivity. 
While these table-top experiments, can provide some insight in EDL dynamics by inferring the motion of tracer particles or using some form of force sensing apparatus~\cite{perkinSternDiffuseLayer2013,tmannetjeElectricallyTunableWetting2013,siretanuDirectObservationIonic2014}, the structure of the EDL has been mostly studied, close to equilibrium, using scattering of high-energy (x-ray or neutron) beams~\cite{penfoldStudiesElectricalDoublelayer1985, boukhalfaSituSmallAngle2014, chuCrowdingAnomalousCapacitance2016} or higher harmonic optical spectroscopy~\cite{lisLiquidFlowSolid2014, debeerSeparatingSurfaceStructure2010}. 
Generally, non-optical methods dominate investigation the EDL dynamics close to surfaces. 
Optical methods, on the other hand, were mainly used for on down-scaling the investigated structures, namely focusing on plasmonic nanoparticles and nanocrystals. 

Studies on plasmonic nanoparticles have mainly attributed the experimental observations to the carrier concentration change in the metallic side and how damping effects, partly caused by surface groups, can alter the localized plasmon resonance linewidth.
Novo and Mulvaney, for example, used dark-field microscope to directly observe the kinetics of deposition onto a single gold nanocrystal and also monitor electro-injection and extraction during oxidation of ascorbic acid on a gold nanocrystal~\cite{novoDirectObservationChemical2008}. 
Later, Collins and Mulvaney optically observed the injection of charges into a single gold nanocrystals in an ion gel environment by using dark field spectroscopy of localized surface plasmon resonance shifts~\cite{collinsSingleGoldNanorod2016}.
Steinhauser et al. have investigated electro-oxidation of a gold nanowire array based on plasmon shifts in acidic electrolytes under different pH values. 
Their observations were in good agreement with cyclic voltametry results~\cite{steinhauserLocalizedPlasmonVoltammetryDetect2018}.
Hoener et al. have reported the shape-dependent spectral response of plasmonic gold nanoparticles~\cite{hoenerSpectralResponsePlasmonic2017}.
In most of these studies, the direct influence of the EDL optical polarizability on the observed signal has been deemed negligible in comparison with the other effect.
In a report focused mainly on the EDL contribution, Dahlin, Zahn, and Vörös concluded the influence of refractive index changes caused by the ions concentration during electric double layer (un)charging are negligible, and almost all of the signal for particles and surface structures that exhibit a localized plasmon resonance can be attributed to the formation of ionic complexes on the metal surface e.g. gold chloride~\cite{dahlinNanoplasmonicSensingMetal2012}.
For a comprehensive overview of research on electrochemical modulation of the optical properties of plasmonic nanoparticles, we recommend reading these recent excellent reviews~\cite{jingNanoscaleElectrochemistryDarkfield2017,hoenerPlasmonicSensingControl2018}.

\subsection{EDL modulation contrast from non-plasmonic structures}
For a metallic particle smaller than 30~nm, where plasmon resonances are suppressed due to high dissipation, or for dielectric particles, the EDL contribution on elastic scattering can reach a fraction of few percents~\cite{naminkElectricDoubleLayerModulationMicroscopy2020,mauranyapinEvanescentSinglemoleculeBiosensing2017a}.
In potentiodynamic measurements of such small particles, this effect can thus become the dominating factor for the observed modulation contrast. 
Previously, we have observed this effect for both metallic and dielectric nanoparticles and have found a reasonable match with the expected value resulting from the EDL modulation~\cite{naminkElectricDoubleLayerModulationMicroscopy2020}. 
Our measurements demonstrated that changing the anion at neutral pH, in the electrochemical potential range dominated by EDL charging, can influence the magnitude of the modulation signal on the same nanoparticle by a factor of two.
That observation can only be explained by the influence of the diffuse layer in the electrolyte on the scattering signal.

In this article, therefore, we focus on restructuring of the EDL when the surface potential is varied, and mainly on the variation of the anion and cation concentrations in the vicinity of the scattering object. 
While the exact optical polarizability of dissolved ions, is generally not so trivial to measure or compute, there is extensive evidence for a linear relationship between concentration of ions and the change in the refractive index.
The observed behavior for ion concentration change as a function of surface charge or size of the scattering object can thus be directly translated to the expected trend for the EDL-modulation contrast as a function of the underlying parameters such as surface charge or local curvature.
Here, we numerically compute the excess ion concentration modulation for three geometries: 1- sphere, 2- cylinder, and 3- hole. 
By investigating the expected sensitivity of this variation to presence of static charges at the interface, we can conclude which of these geometries provide a better chance for observation of single charging events by EDL-modulation microscopy. 

\section{Theory}
The optical contrast mechanism in EDL-modulation microscopy is dependent on the reformation of the screening ions.
The relevant observations timescales in optical experiments are typically much longer the time necessary for the EDL to reach its equilibrium estimated by $\td =  \ld^2/D$, with $D$ the diffusion constant of ions and $\ld$ the Debye length. 
A stationary model of the EDL is therefore sufficient to address the expected signal for realistic experimental conditions.
Another important timescale in this system is the charging time on the electrodes $\tc = L \ld/D$, with $L$ the typical distance between the electrodes~\cite{squiresInducedchargeElectroosmosis2004, bazantDiffusechargeDynamicsElectrochemical2004}. 
For the systems investigated in this article, $\td \approx 10$~ns and $\tc \approx 1$~ms.  
Although optical measurements can be adjusted to investigate either of these timescales, for EDL-modulation microscopy, the modulation frequency can be kept as low as a few Hertz. 
For faster modulation modalities, a time-dependent analysis will become necessary, but that is beyond the scope of the current paper.

\subsection{Modified Poisson-Nernst-Planck}

The traditional EDL theories such as Poisson-Boltzmann (PB) and Poisson-Nernst-Planck (PNP) are based on treating the ions in the electrolyte as point charges.
These methods become invalid for high concentrations and high potentials beyond the thermal voltage of 25 mV since the steric effects become important quantitatively as well as qualitatively~\cite{kilicStericEffectsDynamics2007}. 
Ignoring the steric effects can result in an unphysically high ion concentration in the EDL regime. Consequently, modified PNP (MPNP) equations or classical density functional theories are used to take the finite size of all dissolved species into account~\cite{taoMultiscaleModelingElectrolytes2020,yangChainLengthMatters2020,lianBlessingCurseHow2020}. 
MPNP can self-consistently describe the diffusion and migration of the ions by accounting for steric effects. 
Solving this coupled set of nonlinear equations by advanced large-scale numerical modelling, has been successful in uncovering new phenomena that have not been captured in previous theoretical treatments~\cite{wangSimulationsCyclicVoltammetry2013,liuModifiedPoissonNernst2018,yochelisSpatialStructureElectrical2014}.

The MPNP equations involve the introduction of an effective solvate diameter of all dissolved species. For simplicity, the finite ion sizes are all assumed as $a$ and the maximum local concentration is set to $a^{-3}/N_{A}$. Throughout this work, we set $a = 0.3$~nm, $a^{-3}=61$~M. 

The MPNP equations describe ions flux of each $i$ th ion species $\vec{J}_{i}$ as
%\begin{widetext}
\begin{eqnarray}\label{eq:Nernst-Planck}
%% Eq1
&\vec{J}_{i} = -D_{i} \nabla c_{i}-\frac{D_{i} z_{i} F c_{i}}{R T} \nabla \phi-D_{i} c_{i}\left(\frac{N_{A} \sum_{i=1}^{2} a^{3} \nabla c_{i}}{1- N_{A} \sum_{i=1}^{2} a^{3} c_{i}}\right), \\
%% Eq2
\label{eq:continuity}
&\vec{\nabla} \cdot \vec{J}_{i} = 0,
\end{eqnarray}
%\end{widetext}
where $D_{i}$ is the ion diffusion coefficient, $k_{\mathrm{B}}$ is the Boltzmann constant, $N_{\mathrm{A}}$ is the Avogadro constant, $T$ is the absolute temperature, $F$ is Faraday constant, and ${c_i}$ is net ionic concentration, and $z_i$ is the valence of ionic species. 
Equation ~\ref{eq:continuity} provides the associated continuity equation for each species,~\textit{i.e.}, no ions are created or destroyed.
Additionally, the Poisson equation is used to obtain the mean electrostatic potential  $ \Phi$ from the ion density distributions:
\begin{equation} \label{eq:Poisson}
\vec{\nabla} \cdot ( \epsilon \vec{\nabla} \Phi ) = - e N_{A} (c_{+} - c_{-}),
\end{equation}
with $\epsilon$ the electric permittivity of the medium, and $e$ the elementary charge.

\section{Results}
We investigated three geometries, which encompass a broad range of experimental conditions, for charge sensing based on the EDL-modulation microscopy. 
For each geometry, we look at the variation of the concentration of ions in the EDL as a function of applied potential and then investigate the influence of surface charge and object size on the slope of concentration variation with changing the electrode potentials. 
We focus on the polarizability of EDL and do not include other effects such as change of the plasmonic resonance. 
This estimate can be justified because we look at deeply sub-wavelength metallic or dielectric particles.
Including the effect of charge injection, or chemical reactions at the surface requires a different type of modelling that is beyond the scope of this article.
The EDL effect is present irrespective of the surface reactivity or electrical contact with the particle and hence it deserves an investigation on its own.
Our aim is to understand if the EDL-modulation contrast can reveal the absolute charge of the particle.
For actual experimental conditions, this signal should be separated from the other influential factors by a series of control experiments. 

\begin{figure}
 \centering\includegraphics[width=\textwidth]{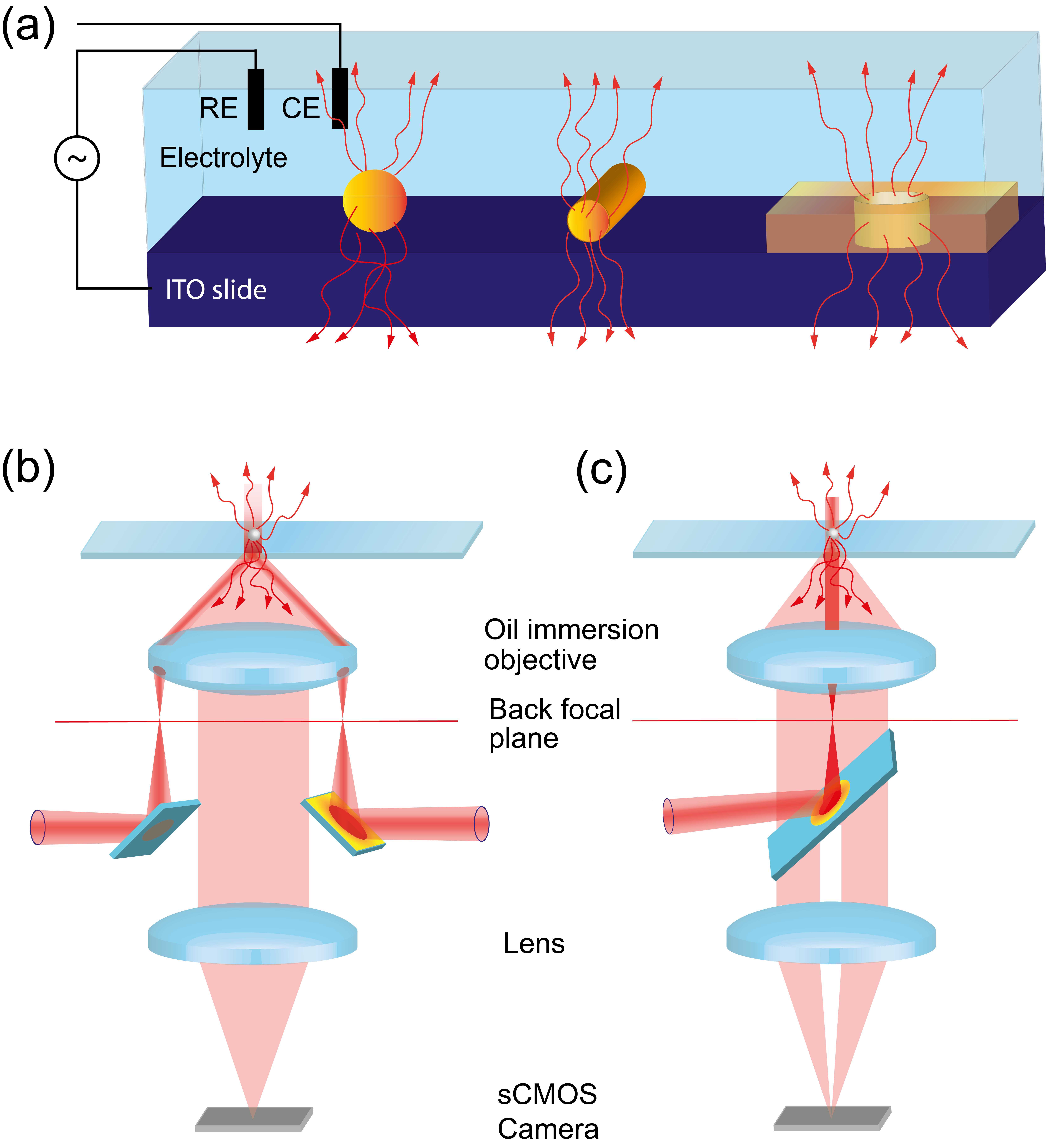}
 \caption{\label{Fig1-setup} (a) Three geometries of sphere, cylinder, and nanohole are considered in this article. The dominating term for EDL-modulation contrast is proportional to the interference of light scattered from the object and the light scattered from the surrounding EDL and can be measured by either of the coherent microscopy techniques such as (b) total internal reflect illumination \cite{naminkElectricDoubleLayerModulationMicroscopy2020,mengMicromirrorTotalInternal2021}  or (c) aperture-shaped interferometric scattering microscopy~\cite{coleLabelFreeSingleMoleculeImaging2017}.}
\end{figure}

The three geometries we consider are schematically depicted in Fig.~\ref{Fig1-setup}(a).
We first consider a nanosphere of given surface charge next to a flat conducting (optically transparent) surface, which plays the role of the working electrode. 
The potential of the flat electrode is controlled relative to the bulk of the liquid. In either total internal reflection scattering or aperture-shaped iSCAT microscopy, depicted in Fig.~\ref{Fig1-setup}(b) and (c), the EDL-modulation signal is determined by the interference of the stationary signal from the particle and scattering from the EDL~\cite{naminkElectricDoubleLayerModulationMicroscopy2020}. As we have discussed previously, this interferometric enhancement is essential for the signal to be detectable within the typical dynamic range of commonly-available scientific cameras.
\subsection{Modelling details}

\begin{figure}
 \centering\includegraphics[width=\textwidth]{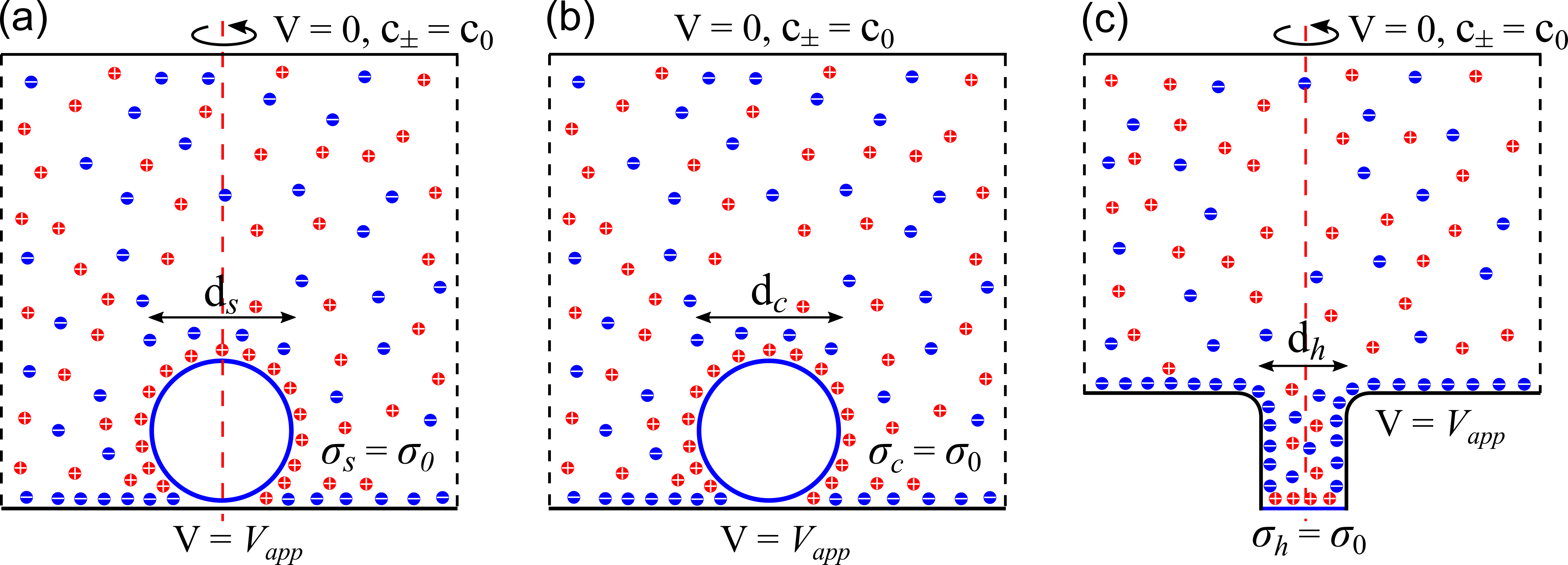}
 \caption{\label{Fig-geo} Geometrical models and set boundary conditions for (a) sphere, (b) cylinder, and (c) nanohole, put on top of conducting electrode and inside an electrolyte solution.}
\end{figure}

The MPNP equations should be solved with the specific boundary conditions. For the initial conditions of the numerical integration, the ion concentrations of each ionic species is assumed to be uniform $c_{\pm}=10 \mathrm{mM}$.
As for the dielectric sphere, cylinder surface, and nanohole bottom surface, these boundary conditions are set: fluxes of the each ionic species are set to zero, $J_{\pm} \cdot \mathrm{n}=0$. The static surface charge densities vary in the range of $\sigma=-(0.01 - 0.12) \mathrm{C} / \mathrm{m}^{2}$, and the electrode potential varies in the range $\psi=(-0.27 - 0.27) \mathrm{V}$.

The boundary conditions on the flat electrode surface are set such that there is no flux of each ionic species, $J_{\pm} \cdot \mathrm{n}=0$, and the constant potential can vary in the range of $\psi=(-0.27 - 0.27) \mathrm{V}$.
The boundary conditions on the top side of the simulation box are constant potential $\psi=0$ and fixed concentration $c_{\pm}=10 \mathrm{mM}$. For the walls depicted with dashed lines in Fig.~\ref{Fig-geo}, the boundary conditions are zero flux for each ionic species and net-charge neutrality.

To capture the details of the double layer, a tailored nonuniform computational mesh is used that is adjusted to the specific requirements for error-free EDL modelling. (i) constant mesh size $0.15$~nm is used near the flat electrode and the sphere/cylinder/nanohole surface with a distant of $0.15$~nm and (ii) uniformly expanding mesh of size $0.15$~nm to size $3$~nm in other area. These meshes are presented in the electronic supplementary information.

\subsection{The sphere}

\begin{figure}[!h]
 \centering\includegraphics[width=\textwidth]{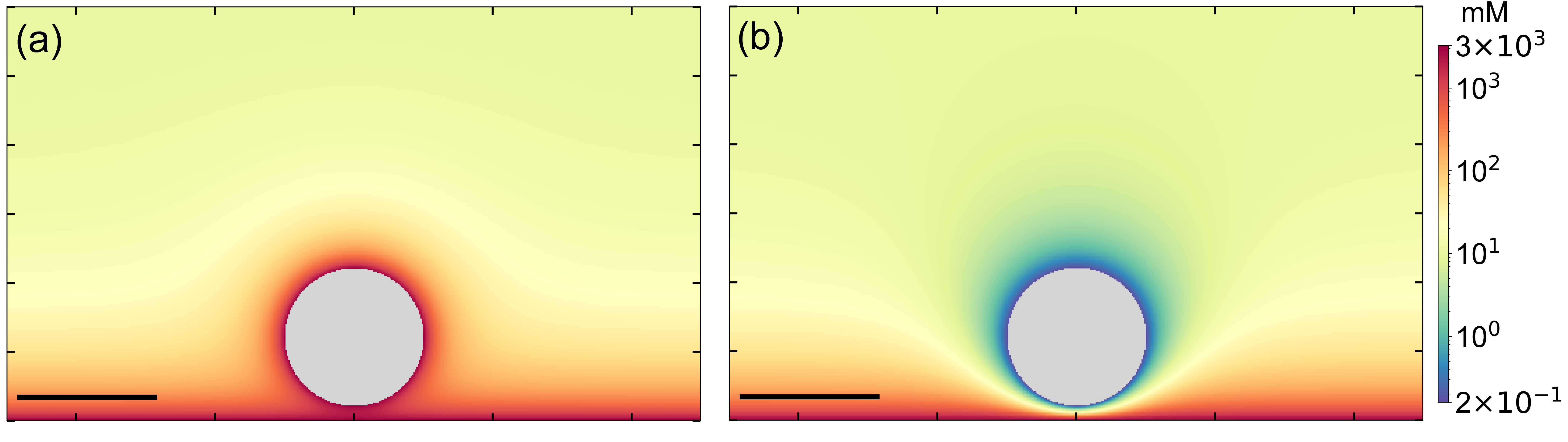}
 \caption{\label{Fig2-sphere_geo} Anion concentration around a sphere lying on a electrode.  The parameters used are  $V_{\mathrm{app}} =  0.15$~volts, $d_{\mathrm{s}} = 5$~nm,  $C_{\mathrm{bulk}} = 10$~mM. The color scale depicts the anion concentration around the sphere for (a) kept at the same surface potential of the electrode and (b) electrically isolated from the electrode, carrying a surface charge of $\sigma_{\mathrm{s}}= -0.02$~C/m$^2$. Scale bars correspond to $5$~nm.}
 \end{figure}

In Fig.~\ref{Fig2-sphere_geo} we plot the calculated ion distribution for a nanosphere on top of an electrode. In our COMSOL model, the sphere and the electrode were immersed in 10~mM KCl aqueous solution. 
Fig.~\ref{Fig2-sphere_geo}(a) depicts the anion concentration in the focal volume with applied potential $V_\mathrm{app}=0.15$~volts for a conducting sphere that is electrically connected to the electrode. 
Fig.~\ref{Fig2-sphere_geo}(b) presents the anion concentration in the same area with applied potential $V_\mathrm{app}=0.15$~volts for a dielectric sphere that is electrically separated from the electrode, with a surface charge on the sphere of $\sigma_{\mathrm{s}}= -0.02$~C/m$^2$. 
This numerical calculation is repeated for a range of surface charge values and sphere sizes. 

\begin{figure}[!h]
 \centering\includegraphics[width=\textwidth]{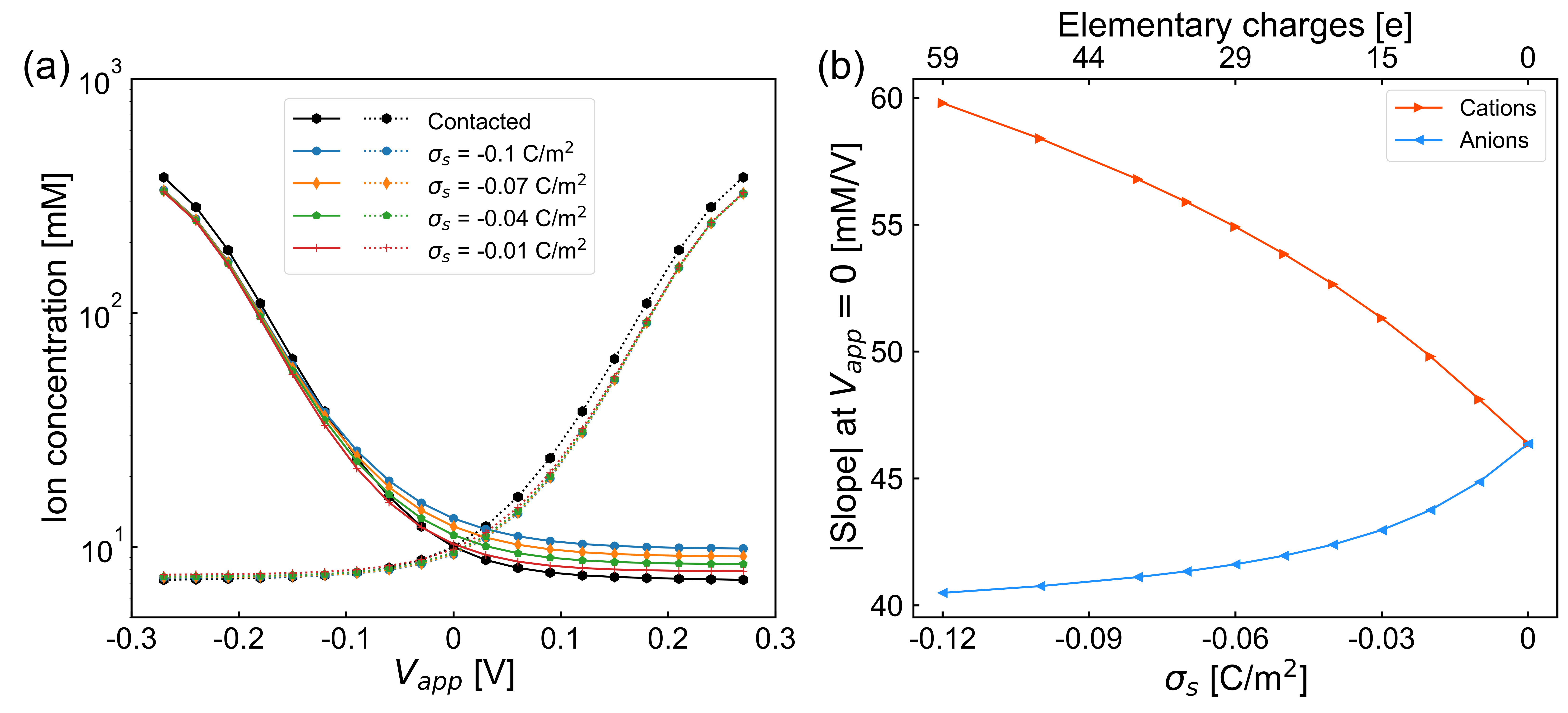}
 \caption{\label{Fig2-sphere_surfacecharge} (a) Ion concentration in the focal volume, for different values of the applied potential. Each curve corresponds to a different value of the surface charge on the sphere. For comparison, the same quantity is computed for a conducting sphere raised to the same potential as the electrode. The sphere diameter is  $d_{\mathrm{s}} = 5$~nm. (b) Absolute slope of the curves in (a) at $V_{\mathrm{app}} = 0$~volts for various values of charge on the sphere.}
\end{figure}

In Fig.~\ref{Fig2-sphere_surfacecharge}, we plot the change of the anion concentration (dashed curves) and cation concentration (solid curves) in the depicted volume, as a function of the electrode potential.
In Fig.~\ref{Fig2-sphere_surfacecharge}(a), each curve corresponds to a different value of the total charge of the sphere. At a relatively large negative potential on the electrode, the cation and anion concentrations are almost equal for different surface charge on the sphere. As the applied potential increases from $-0.1$ to $0.27$~volts, the cation concentration exponentially decreases and the concentration around the sphere remains larger for higher surface charge values. 
For anion concentration, with increasing applied potential, we observe an almost exponential increase that levels at positive surface potentials, independent of the surface charge on the sphere. 
This increase starts to level off at a potential of close to 0.3 volts.
For comparison, we plot the expected change in the anion concentration for a conducting particle that is kept at the same potential of the electrode.

In EDL-modulation microscopy, the change of the scattering intensity is dominated by the ion species with the higher optical polarizability. 
For alkali metal halide solutions, this signal is mainly proportional to the anion concentration because of the larger polarizability of halogen ions. 
From refractive index tables, it seems that generally the alkali cations contribute less to variations of the refractive index, but it is easily possible to include their effect with our method.
In Fig.~\ref{Fig2-sphere_surfacecharge}(b), we plot the absolute value of the slope of ion concentration with change of applied potential around $V_{\mathrm{app}} = 0$~volts for the curves in Fig.~\ref{Fig2-sphere_surfacecharge}(a). 
The slope of cation concentration decreases from $60$~mM/V to $47$~mM/V as the surface charge on sphere changes from $-0.12$~C/m$^2$ to $0$~C/m$^2$, corresponding to exchanging roughly 60 elementary charges. 
In contrast, the slope of anion concentration slightly increases from  $40$~mM/V to $47$~mM/V.
At an electrode potential of zero volts, the cations and anions are the same in the depicted volume as for the contacted sphere, while the density of counter-ions becomes higher than that of co-ions with larger negative surface charge to compensate the negative charge of sphere.
It can be seen that first derivatives of both ionic concentrations at 0 volts are the same for an equipotential sphere. 
However, the absolute slope of cations increases with the surface charge density of spheres, while it is the reverse trend for the anions. 
This observation suggests that cations become more sensitive to the external voltage as more cations are located near the negative charge sphere. 
The difference between change of anion and cation concentrations indicates that the total ion density will increase in the depicted volume in response to varying the electrode potential.

We observe that at the large negative voltage, the cation concentration for contacted sphere is even larger than an electrically-isolated negatively charged spheres. For a sphere with surface charge -0.1 C/m$^2$, the equivalent zeta potential on the surface of sphere is about $-0.116$~volts. Therefore, decreasing the sphere potential to negative values beyond that will attract a denser layer of cations around the sphere.

\begin{figure}[!h]
 \centering\includegraphics[width=\textwidth]{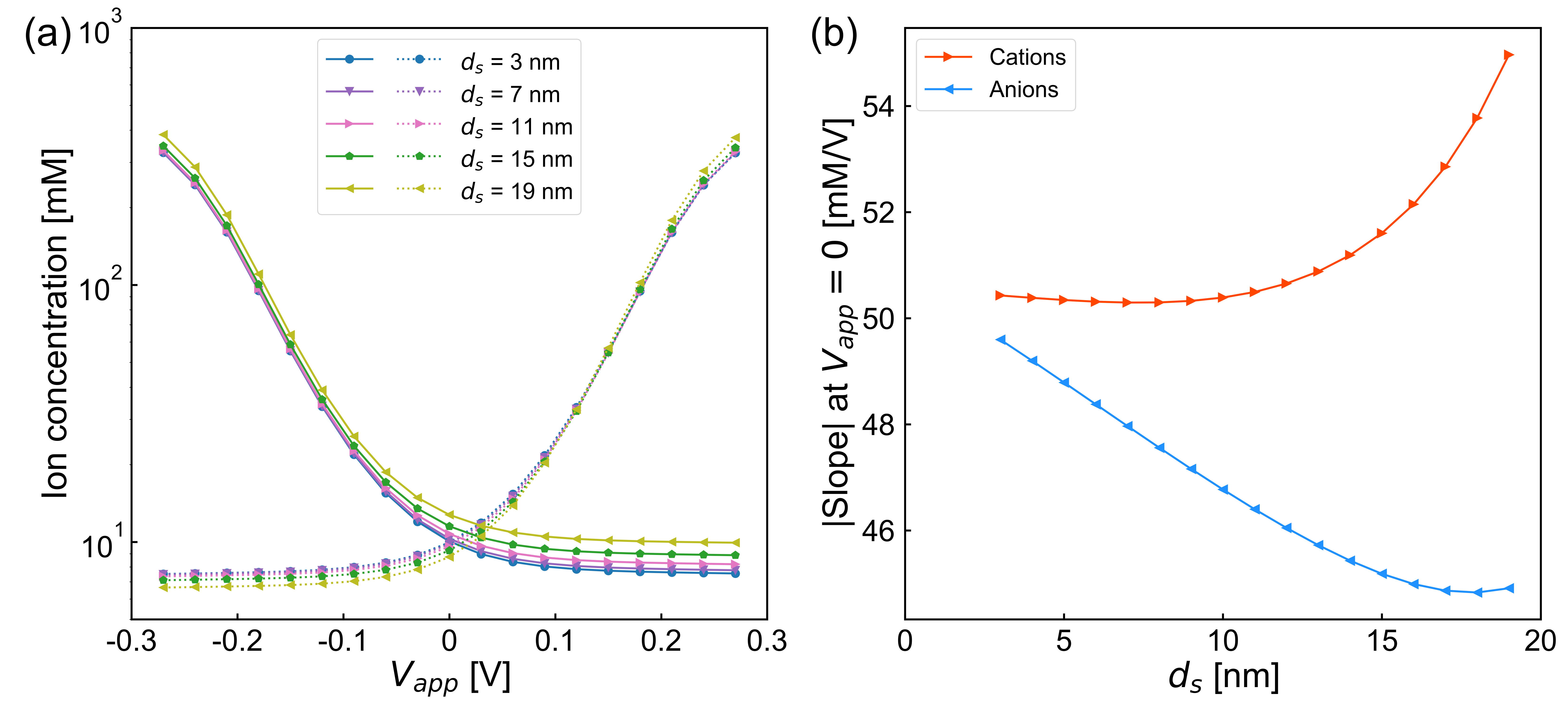}
 \caption{\label{Fig2-sphere_radius} (a) Ion concentration in the focal volume, for different size of the sphere. Each curve correspond to a different size of the sphere. Here the surface charge of the sphere is   $\sigma_{\mathrm{s}}= -0.02$~C/m$^2$. (b) Absolute slope of the curves in (a) at $V_{\mathrm{app}} = 0$~volts for various values of the sphere diameter.}
\end{figure}

Next, we check the ions concentration in the focal volume around spheres of different size. 
For this part, the sphere has a dielectric constant of $\epsilon = 4$ and the surface charge of sphere is fixed at $\sigma_{\mathrm{s}}= -0.02$~C/m$^2$.
In Fig.~\ref{Fig2-sphere_radius}(a) each curve corresponds to a different diameter of the sphere, the cation(anion) concentration decrease(increase) almost exponentially with the applied potential at the electrode.
Cation concentration around the sphere with a larger diameter is higher than that around the smaller sphere. 
The balance between the packing of the ions to compensate the surface charge and the interactions between the ions leads to the variation in the density profiles of ions for spheres with different diameters. 
The total charge of ions, i.e. the total ionic charge within a distance from the charged spheres for fixed surface charge density, scales quadratically at zero voltage with the radius of the sphere. 
Therefore, the spheres at fixed surface charge density with various diameter exhibit an increase in counter-ions (cations), while the co-ions (anions) decrease due to the steric effects~\cite{yu_density-functional_2004,yang_curvature_2020}. 
A similar behavior is expected for cylinder ~\cite{huang_curvature_2010} and nanohole geometries.
In Fig.~\ref{Fig2-sphere_radius}(b), we plot the absolute slope of ion concentration with applied potential around $V_{\mathrm{app}} = 0$~volts for the curves in Fig.~\ref{Fig2-sphere_radius}(a). 
The slope of cations concentration first remain constant around $50$~mM/V as the diameter of the sphere $d_{\mathrm{s}}$ is raised from 3~nm to 10nm, and then rapidly increases with further increase of sphere diameter. 
The slope of anion concentration decreases from $50$~mM/V and then plateaus at $45$~mM/V while the diameter is changed from $3$~nm to $20$~nm. 
Consequently, the total charge variation in responds to the external voltage in the depicted volume first decreases, and then increases with the increase of sphere diameter. 

\subsection{The cylinder}

\begin{figure}
 \centering\includegraphics[width=\textwidth]{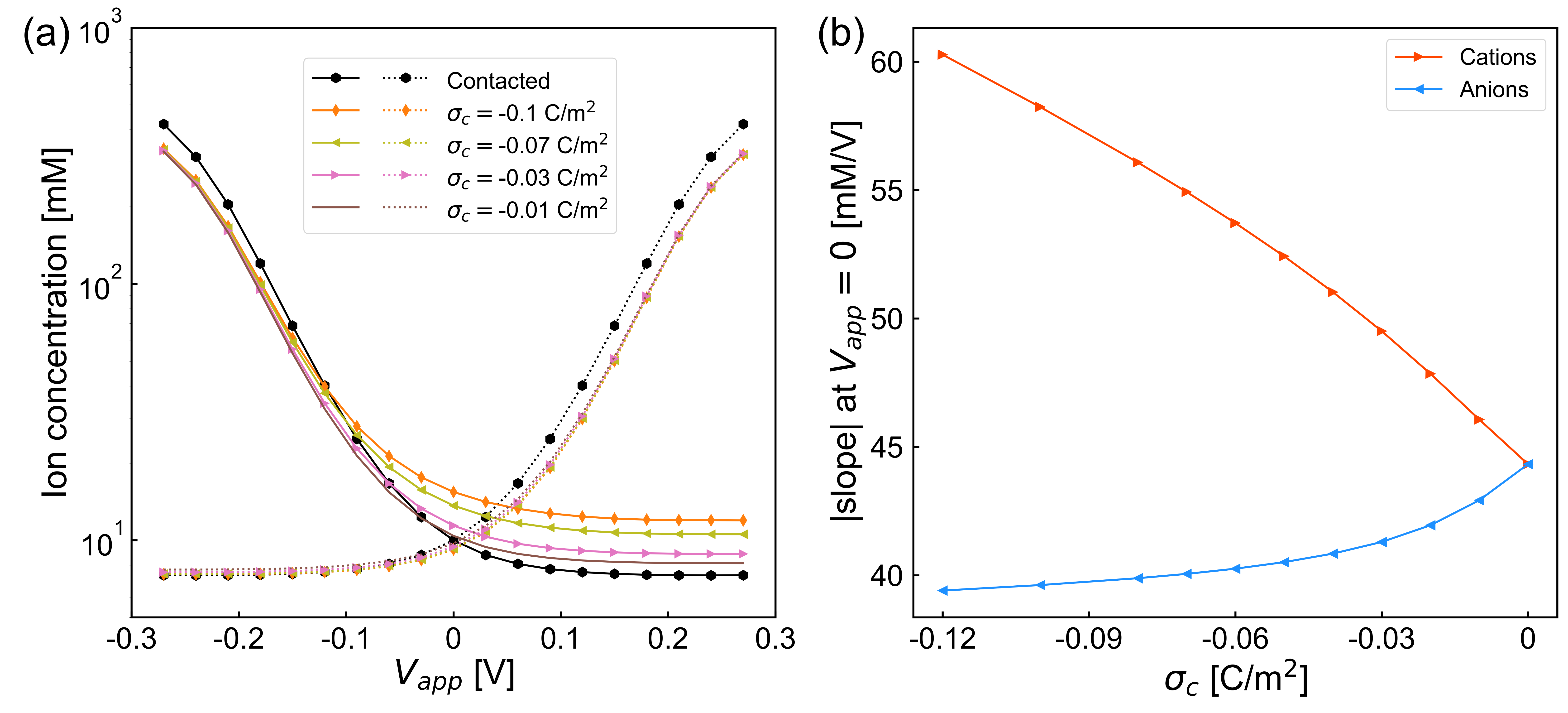}
 \caption{\label{Fig3-cylinder_surfacecharge} (a) Ion concentration in the focal volume around Cylinder, relative to the bulk, for different values of the applied potential. Each curve correspond to a different value of the surface charge on the cylinder. For comparison, the same quantity is computed for a conducting sphere raised to the same potential as the electrode. The cylinder diameter is  $d_{\mathrm{c}} = 5$~nm. (b) Absolute slope of the curves in (a) at point of  applied potential $V_{\mathrm{app}} = 0$~volts for various values of charge on the sphere.}
\end{figure}

Next we consider a cylinder instead of a sphere next to the electrode.
In experiments, this geometry corresponds to investigating a nanowire placed on the working electrode.
The results are depicted in Fig.~\ref{Fig3-cylinder_surfacecharge}, obtained following the same procedure as for the sphere. 
The response of the ion concentration to the electrode potential is generally similar to that of a sphere, even though the total charge in the EDL scales linearly with radius of the cylinder. 

\begin{figure}
 \centering\includegraphics[width=\textwidth]{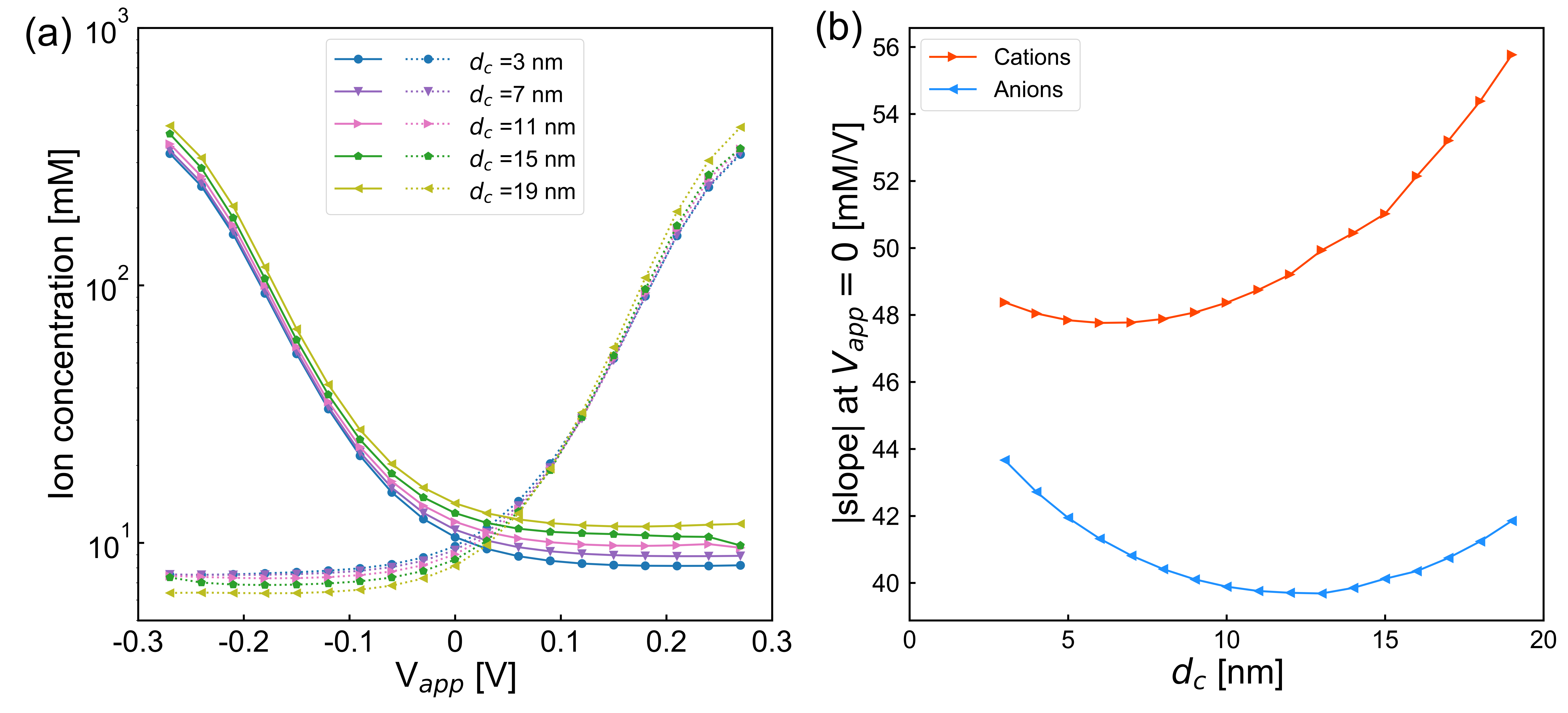}
 \caption{\label{Fig3-cylinder_radius}(a) Ion concentration in the focal volume around the cylinder, for different sizes of the cylinder at a surface charge $\sigma_{\mathrm{c}}= -0.02$~C/m$^2$. (b) the absolute slope of the curves in (a) at point of applied potential $V_{\mathrm{app}} = 0$~volts for various diameters.}
\end{figure}

Similarly, we check the ions concentration in the focal volume around cylinders of varying.
The cylinder is electrically separated with the electrode and the surface charge of cylinder is fixed at $\sigma_{\mathrm{c}}= -0.02$~C/m$^2$.
Fig.~\ref{Fig3-cylinder_radius}(a) shows the same trend of the ions concentration changes with applied potential on electrode. 
In Fig.~\ref{Fig3-cylinder_radius}(b), we calculated the absolute slope of ion concentration with applied potential around $V_{\mathrm{app}} = 0$~volts for the curves in Fig.~\ref{Fig3-cylinder_radius}(a). 
The slope of the cation and anion concentration curves have similar parabolic shape, with the diameter of the cylinder increasing from $3$~nm to $20$~nm, the slope decreases quadratically at first and then increases for larger diameters.
However the slopes reach their minima at different diameters for cations and anions. 
For cations the lowest absolute slope is at $d_\mathrm{c}= 7.5$~nm, while the for the anions is at $d_\mathrm{c}= 12.5$~nm.
Although the surface area for cylinder scales differently with diameter than that of the sphere, the dependence of the concentration variation with potential on diameter is qualitatively the same for both geometries.

\subsection{The nanohole}

\begin{figure}
 \centering\includegraphics[width=\textwidth]{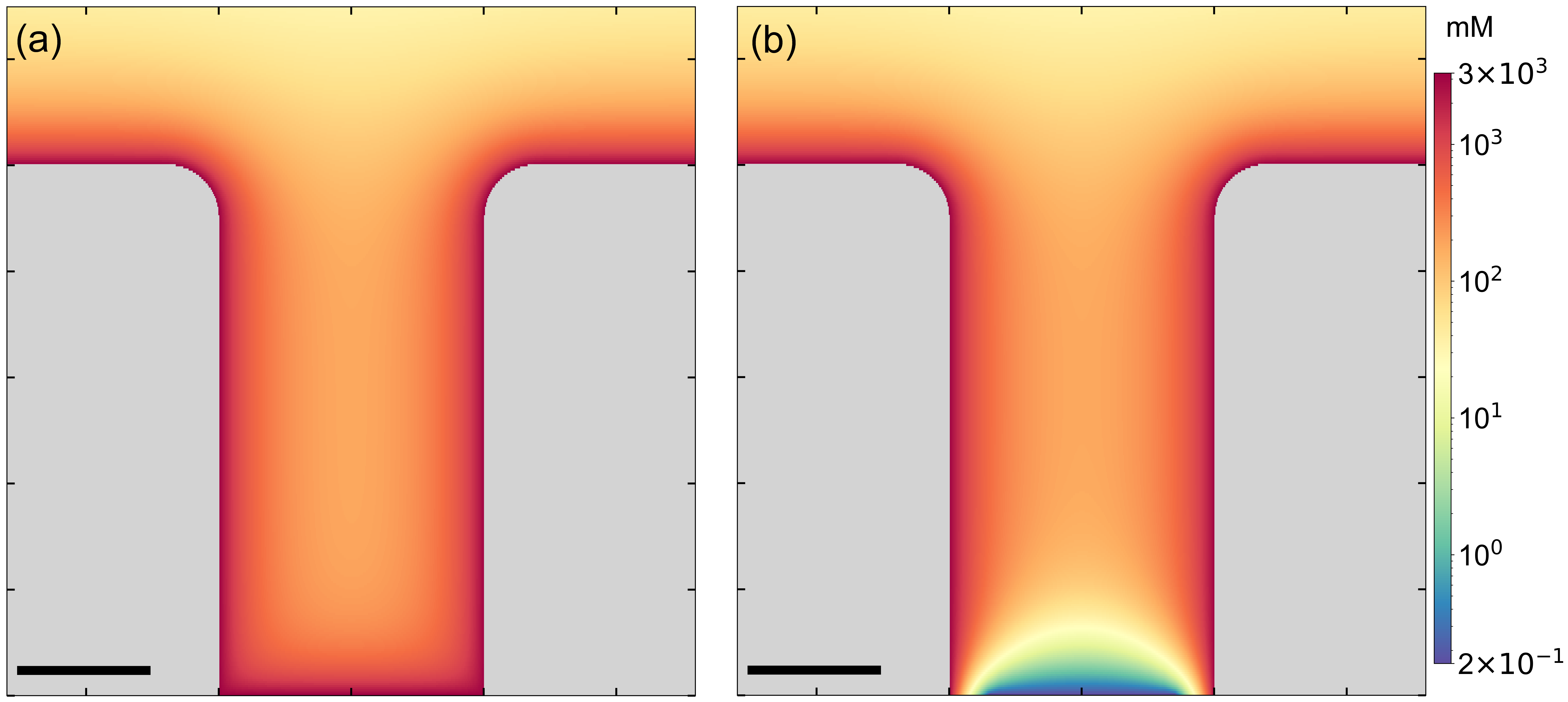}
\caption{\label{Fig4-hole_geo} Anion concentration inside and surrounding a nanohole. The parameters used are  $V_{\mathrm{app}} = 0.15$~volts, $d_{\mathrm{h}} = 5$~nm, $H_{\mathrm{h}} = 10$~nm, $C_{\mathrm{bulk}} = 10$~mM.(a) The anion concentration inside of the nanohole which is electrically connected to the bottom surface. (b) The anion concentration inside of the nanohole which is electrically separated with bottom surface, the surface charge on the bottom surface is $\sigma_{\mathrm{b}}= -0.02$~C/m$^2$. Scale bars are $2.5~$nm}
\end{figure}

The last geometry we consider is a nanohole drilled inside the electrode as in Fig.~\ref{Fig4-hole_geo}.
Here the walls are at the same potential of the electrode, while the bottom of the hole has a constant surface charge.
This geometry is most representative of top-down nanofabrication on flat electrodes, and especially important when considering its connection with nanopores used in Coulter-type sensing of biomolecules.
The charge, size, and potential parameters in this model are in the same range as for the sphere and the cylinder and the electrolyte is 10~mM KCl aqueous solution. 
Fig.~\ref{Fig4-hole_geo}(a) shows the anion concentrations in the focal volume inside of the nanohole with applied potential $V_\mathrm{app}=0.15$~volts when considering the bottom of the nanohole to be electrically connected to the electrode. Fig.~\ref{Fig4-hole_geo}(b) shows the anion concentration in the same area with applied potential $V_\mathrm{app}=0.15$~volts for a charged bottom of surface density $\sigma_{\mathrm{b}}= -0.02$~C/m$^2$.

\begin{figure}
\centering\includegraphics[width=\textwidth]{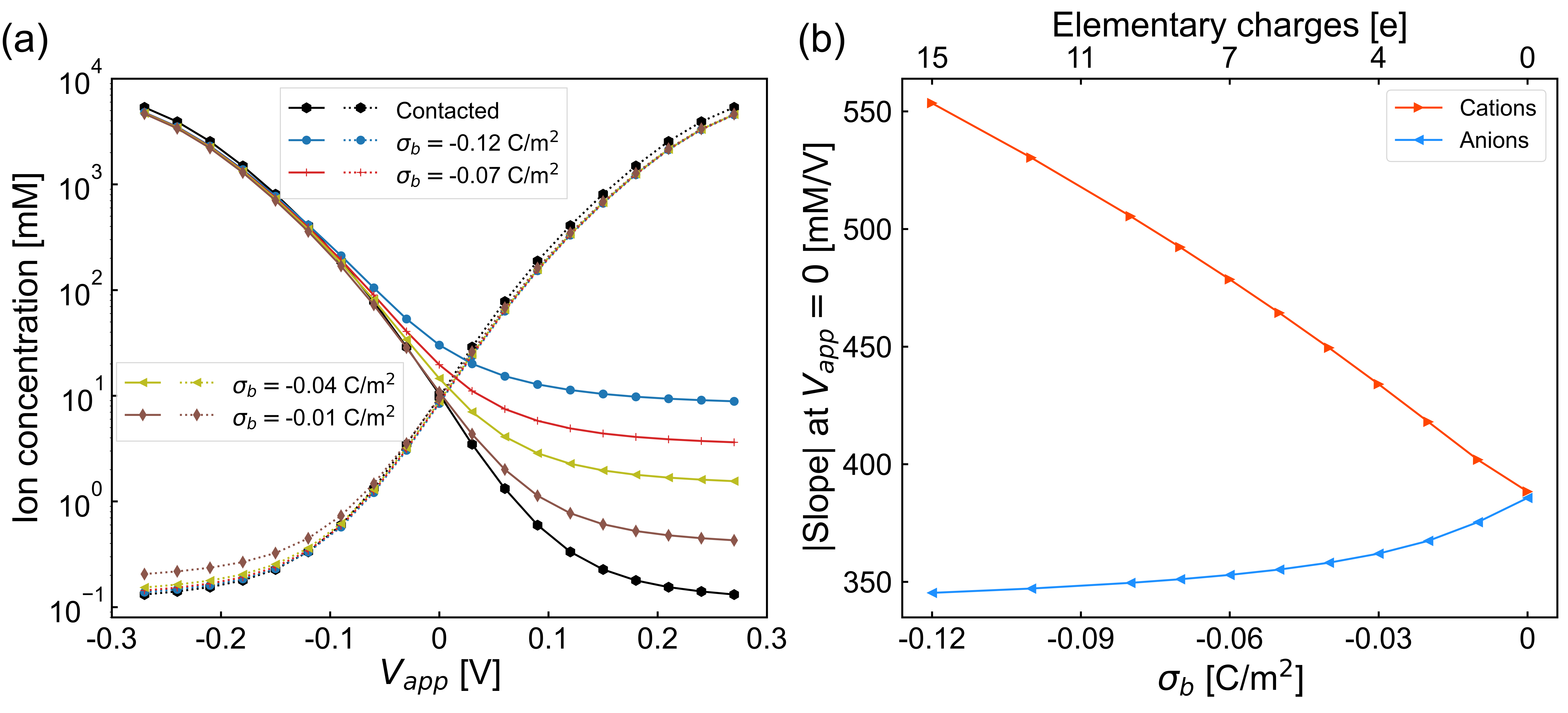}
\caption{\label{Fig4-hole_surfacecharge}(a) Ion concentration inside the nanohole, for different values of the applied potential. Each curve correspond to a different value of the surface charge on the bottom surface. For comparison, the same quantity is computed for a conducting bottom surface raised to the same potential as the walls. The nanohole diameter is  $d_{\mathrm{h}} = 5$~nm, the height of the nanohole is $H_{\mathrm{h}} = 10$~nm. (b) Absolute slope of the curves in (a) at $V_{\mathrm{app}} = 0$~volts for various values of charge on the bottom surface. }
\end{figure}

We plot the modelling results for the nanohole in Fig.~\ref{Fig4-hole_surfacecharge}, following the same routine of the sphere and cylinder described before. 
In this case, the nanohole walls are kept at the same potential of the electrode. 
In Fig.~\ref{Fig4-hole_surfacecharge}(a), each curve corresponds to a different value of the total charge on bottom surface of the nanohole. 
In presence of a large negative potential on the electrode, both cation and anion concentrations are almost at the same level for different values of surface charge on the bottom surface. 
As the applied potential is increased from $-0.1$ to $0.27$~volts, cation concentrations decrease exponentially reaching a limiting values that depends on the surface charge at the bottom of the whole. 
For anions, at large positive potential, the concentration reaches the same level almost independent of the surface charge at the bottom. 
The difference between the cations concentration for different surface charge on the bottom in this potential window is much larger than that of the sphere. 
In Fig.~\ref{Fig4-hole_surfacecharge}(b), we plot the absolute slope of ion concentration dependence on electrode potential at $V_{\mathrm{app}} = 0$~volts for the curves in Fig.~\ref{Fig4-hole_surfacecharge}(a). 
The absolute value of the slope of cation concentration decreases from $550$~mM/V to $390$~mM/V with the surface charge on bottom surface increase from $-0.12$~C/m$^2$ to $0 $~C/m$^2$, corresponding to a change of roughly 15 elementary charges. 
The slope of anion concentration varies relatively less, increasing from  $350$~mM/V to $390$~mM/V. 
Comparing Fig.~\ref{Fig4-hole_surfacecharge} with Fig.~\ref{Fig2-sphere_surfacecharge}, the curves of ions concentration with the change of applied potential and the slope of the curve around the applied potential $V_{\mathrm{app}}=0$~volts are similar and the trend and shape are also similar, however the change of the ions concentration and the slopes values of nanohole are 10 times lager than the corresponding values for the sphere.
This difference can be attributed to the confinement of the observation volume to the nanohole walls, which is also the main contribution to the change in the scattering signal.
The effective observation volume for sphere is, in contrast, defined by the optical point spread function. 

\begin{figure}
\centering\includegraphics[width=\textwidth]{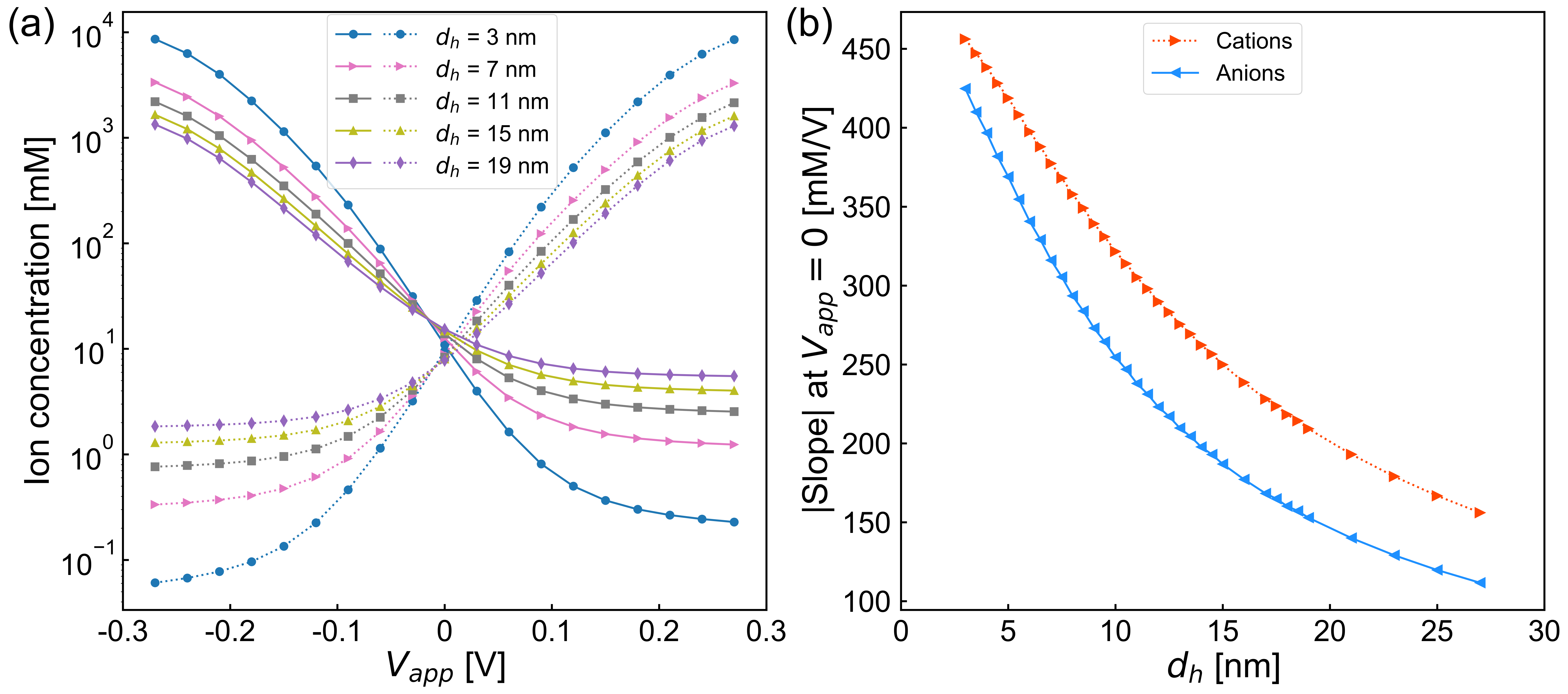}
 \caption{\label{Fig4-hole_radius} (a) Average ion concentration inside the nanohole, for different size of the nanohole. Each curve corresponds to a different size of the nanohole. For this plot, the surface charge of the bottom is $\sigma_{\mathrm{b}}= -0.02$~C/m$^2$. (b) Absolute slope value of the curves in (a) at point of applied potential $V_{\mathrm{app}} = 0$~volts for various values of nanohole diameters.}
\end{figure}

Similar to Fig.~\ref{Fig2-sphere_radius} and Fig.~\ref{Fig3-cylinder_radius}, in Fig.~\ref{Fig4-hole_radius} we compare the behavior of ion concentrations inside nanoholes of different sizes. 
The surface charge at the nanohole bottom is fixed at $\sigma_{\mathrm{b}}= -0.02$~C/m$^2$.
In Fig.~\ref{Fig4-hole_radius}(a) the cation concentration inside of the larger nanohole is lower than that inside of smaller nanohole when the applied potential increasing from $-0.27$ to $0$~volts, and the cation concentration difference between each curves for different size of nanohole gradually decrease to 0 until the applied potential $V_{\mathrm{app}}=0$~volts and then the difference start to increase with the increasing of the applied potential from $0$ to $0.27$~volts.
Cation concentration inside of a larger nanohole is higher than that insdie of smaller nanohole in the range of applied potential from $0$ to $0.27$~volts.
The trends and shapes of the curves for ions concentration is just opposite to that of cations concentration.
In Fig.~\ref{Fig4-hole_radius}(b), we calculated the absolute slope of ion concentration with change of applied potential around $V_{\mathrm{app}} = 0$~volts for the curves in Fig.~\ref{Fig4-hole_radius}(a).
The absolute value of slopes of cation and anion concentration curves have similar trend. 
These slopes decrease as the diameter of the nanohole increases from $3$~nm to $28$~nm. 
The slopes of cations concentration curves is a few percent larger than the slope of the anions concentration curve.

\section{Discussion}
Microscopy methods based on (phase-modulated) linear scattering have attracted a lot of interest in recent years because of their remarkable sensitivity, which is even sufficient for detecting single macromolecules. 
A wide range of these methods, however, are still dependent on differential measurement to detect such small features. 
In this article, we have demonstrated using numerical modeling that the EDL-modulation contrast is sensitive to both local curvature and local surface charge.
This sensitivity, enables continuous investigation of stationary objects on the surface, by altering the electrode potential, instead of relying on observation of landing objects. 
Interestingly, this contrast increases for smaller objects as the polarisability of dielectric scatterers scale with their volume, while the EDL contribution to scattering generally scales with the surface area. 
Considering the optical diffraction limit, flatness of the substrate and stability of the background also play major roles in the feasibility of detecting the EDL-contrast from small objects.
Drilling nanoholes in an opaque substrate is a relatively simple but effective way of separating signal of the target molecules from the bulk fluctuations. 
In florescence-based single molecule detection, researchers have been using this solution for almost two decades, calling the nanoholes zero-mode waveguides ~\cite{leveneZeroModeWaveguidesSingleMolecule2003}.

Our calculations on the three common detection geometries demonstrate that nanoholes can provide an EDL-modulation contrast that is highly sensitive to the static charge at the bottom of the nanohole. 
Most remarkably, sensitivity to hopping of a single elementary charge, or equivalent to that, is within technical reach of interferometric methods.
Reaching this ultimate sensitivity in a completely label-free method can be a breakthrough in probing the intermediate steps of complex chemical reactions.
One of the major directions will be studying biomolecular interactions, where the complexity of processes and large number of involved compounds make single molecule assays, a dominant choice in many laboratories and for some investigations an inevitable choice. 
As discussed in the introductory remarks, separating the EDL-modulation scattering signal from that of forming chemical compounds at the interface will remain a technical challenge. 
Fortunately, this microscopy method provides several controlling parameters such as the electrochemical window and modulation frequency to separate the EDL effect from other surface processes in the vast temporal-electrochemical phase space.  

To overcome the limitations of optical diffraction, while taking advantage of the speed and simplicity of this technique, we foresee that several application of EDL-modulation microscopy can emerge in combination of bottom-up nanofabrication, using for example carbon nanotubes as electrodes. 
Recent observations of optical scattering signals from charging of two-dimensional semiconductor materials~\cite{zhuOpticalImagingCharges2019} and injection of lithium into in battery electrode materials~\cite{merryweatherOperandoOpticalTracking2020} indicate that using potentiodynamic modulation of the local refractive index as an optical contrast is forming a new frontiers for exploration at the interface of two very strong analytical disciplines, optical microscopy and electrochemistry.

% \begin{table}
% \end{tabular}
% \end{table}

\section{CRediT author statement}
\textbf{Zhu Zhang}: Visualization, Investigation, Writing- Original draft preparation.
\textbf{Jie Yang}: Methodology, Software, Validation, Writing- Original draft preparation.
\textbf{Cheng Lian}: Methodology, Validation, Writing- Reviewing and Editing, Supervision.
\textbf{Sanli Faez}: Conceptualization, Methodology, Writing- Original draft preparation, Writing- Reviewing and Editing, Supervision.

\begin{acknowledgments}
We thank Allard P. Mosk and Serge Lemay for fruitful discussions.
This research was supported by the Netherlands Organization for Scientific Research (NWO grant 680.91.16.03), China Scholarship Council (CSC 201806890015), and National  Natural  Science Foundation  of  China  (22078088).
\end{acknowledgments}

\bibliography{JAPBib}
\end{document}